\documentclass[reprint, aps, superscriptaddress]{revtex4-2}

\usepackage{float}
\usepackage{graphicx}
\usepackage{dcolumn}
\usepackage{bm}
\usepackage{amsmath}
\usepackage[title]{appendix}

\begin{document}

\title{Topological Robustness of Anyon Tunneling at $\nu = 1/3$}

\author{Adithya Suresh}
\affiliation{Department of Physics and Astronomy, Purdue University, West Lafayette, IN, 47907}
\author{Ramon Guerrero-Suarez}
\affiliation{Department of Physics and Astronomy, Purdue University, West Lafayette, IN, 47907}
\author{Tanmay Maiti}
\affiliation{Department of Physics and Astronomy, Purdue University, West Lafayette, IN, 47907}
\author{Shuang Liang}
\affiliation{Department of Physics and Astronomy, Purdue University, West Lafayette, IN, 47907}
\author{Geoffrey Gardner}
\affiliation{Microsoft Quantum, West Lafayette, IN, 47907}
\author{Claudio Chamon}
\affiliation{Department of Physics and Astronomy, Purdue University, West Lafayette, IN, 47907}
\affiliation{Purdue Quantum Science and Engineering Institute, Purdue University, West Lafayette, IN, 47907}
\author{Michael Manfra}
\email{mmanfra@purdue.edu}
\affiliation{Department of Physics and Astronomy, Purdue University, West Lafayette, IN, 47907}
\affiliation{Microsoft Quantum, West Lafayette, IN, 47907}
\affiliation{Purdue Quantum Science and Engineering Institute, Purdue University, West Lafayette, IN, 47907}
\affiliation{Elmore Family School of Electrical and Computer Engineering, Purdue University, West Lafayette, IN, 47907}
\affiliation{School of Materials Engineering, Purdue University, West Lafayette, IN, 47907}

\date{\today}

\begin{abstract}
The scaling exponent $g$ of the quasiparticle propagator for incompressible fractional quantum Hall states in the Laughlin sequence is expected to be robust against perturbations that do not close the gap. Here we probe the topological robustness of the chiral Luttinger liquid at the boundary of the $\nu=1/3$ state by measuring the tunneling conductance between counterpropagating edge modes as a function of quantum point contact transmission. We demonstrate that for transmission $t\geq 0.7$ the tunneling conductance is well-described by the first two terms of a perturbative series expansion corresponding to $g=1/3$. We further demonstrate that the measured scaling exponent is robustly pinned to $g=1/3$ across the plateau, only deviating as the bulk state becomes compressible. Finally we examine the impact of weak disorder on the scaling exponent, finding it insensitive. These measurements firmly establish the topological robustness of anyon tunneling at $\nu=1/3$ and substantiate the chiral Luttinger liquid description of the edge mode.
\end{abstract}

\maketitle

The edge modes circulating at the boundary of a fractional quantum Hall effect (FQHE) state are described by chiral Luttinger liquid theory which encodes the correspondence to the underlying topological order in the bulk~\cite{Wen1990, Blok1990}. For Laughlin states at filling fraction $\nu=1/m$, the scaling exponent is given by $g = \nu$ for tunneling of fractionally charged quasiparticles (anyons)~\cite{Wen1990, Wen1991, Wen1995, Kane1992, Kane1992PRL, Kane1994, Fendley1994, Fendley1995PRL, Fendley1995PRB, Chamon1995, ChamonWen1992}. Experimentally, the scaling exponent can be extracted from analysis of the tunneling conductance across a quantum point contact (QPC) \cite{Roddaro2003, Roddaro2004PRL, Roddaro2004SolidState, Baer2014, Ensslin2018, guerrerosuarez2025universalanyontunnelingchiral, Cohen2023} or by examination of current fluctuations (noise) at the thermal-to-shot noise crossover \cite{veillon2024observation}. When combined with measurements of the effective anyon charge $e^*$ and the anyonic statistical angle $\theta_a$, determination of $g$ completely specifies the topological order of the ground state responsible for quantization in the bulk, connecting the physics at the edge to the properties of the bulk quantum Hall state.  

\begin{figure*}[!htb]
    \includegraphics[width=\textwidth]{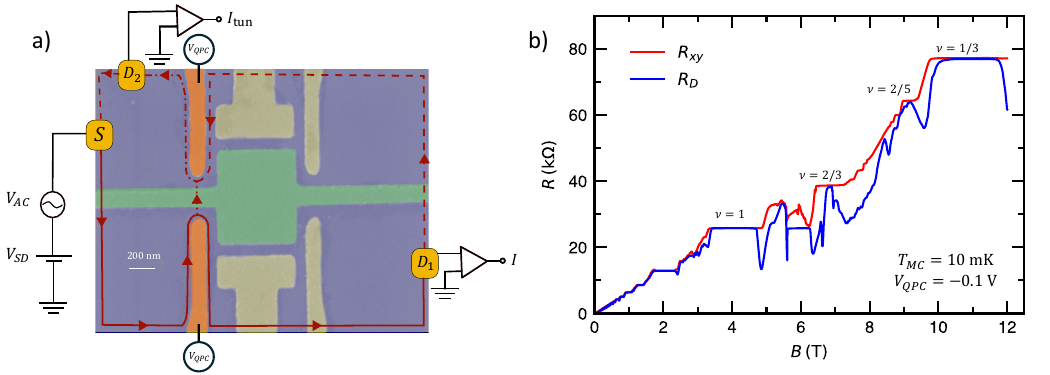}
    \caption{\textbf{(a)} False color scanning electron microscope image of a device similar to the one measured in this experiment including a schematic of the measurement circuit. The QPC used for tunneling experiments is highlighted in orange and the helper gate is highlighted in green. The red lines show circulation of the chiral edge modes. The solid line corresponds to the edge mode equilibrated with the source contact and carries the non-equilibrium current. Dashed lines are edge modes equilibrated with the drain contacts and the dotted line corresponds to the tunneling current across the QPC. $S$, $D_1$ and $D_2$ are the source and drain contacts. \textbf{(b)} Simultaneous measurement of the Hall resistance across the bulk ($R_{xy}$) and across the QPC ($R_D$) as a function of magnetic field at a mixing chamber temperature of $T_{\text{MC}} = 10 \text{mK}$. The QPC is biased just past depletion to define the current path while still fully transmitting the edge modes. The filling fraction in the QPC is the same as in the bulk.}
    \label{Fig1}
\end{figure*}

The edge mode at $\nu = 1/3$ was first described as a chiral Luttinger liquid (CLL) by Wen \cite{Wen1990}; the tunneling conductance between counterpropagating modes at weak backscattering was derived using perturbation theory \cite{Wen1991}. Following these theoretical advances, several experimental attempts were undertaken to test the predicted scaling behavior for quasiparticle tunneling. These early experiments resulted in significant discrepancies with the predictions for the chiral Luttinger liquid, both qualitatively and quantitatively \cite{Milliken1996SSC, Roddaro2003, Roddaro2004PRL, Roddaro2004SolidState, Baer2014, Ensslin2018}. It is possible that these discrepancies were due in part to edge reconstruction due to the combination of soft confinement potential and disorder \cite{ChamonWen1994, Umansky2019}. Additional experiments were also in contradiction with theoretical expectations. For example, experiments using cleaved edge overgrowth heterostructures \cite{Chang1996PRL, Grayson1998PRL, Chang2001PRL} investigated tunneling of electrons from a 3D gas tunnel-coupled to a fractional quantum Hall edge. The authors of this work studied the filling factor dependence of the scaling exponent, but found that $g$ varied continuously across the $\nu=1/3$ plateau whereas quantization within the plateau is expected from theory. 

In a recent experiment using an AlGaAs/GaAs heterostructure designed to have sharp confinement \cite{guerrerosuarez2025universalanyontunnelingchiral}, we demonstrated scaling behavior for tunneling of anyons in a QPC for $\nu=1/3$. The tunneling conductance, both qualitatively and quantitatively, exhibited scaling behavior consistent with chiral Luttinger liquid predictions. A recent experiment utilizing a graphene device saw similar success in demonstrating consistent CLL behavior for the tunneling of electrons between $\nu=1$ and $\nu=1/3$ edge modes \cite{Cohen2023}. 

 Here we report on experiments designed to probe the topological robustness of the chiral Luttinger liquid scaling exponent at $\nu= 1/3$. We present a detailed study of the tunneling conductance as a function of QPC transmission. Our analysis establishes the range of transmission over which $g=1/3$ and perturbative calculations in the weak backscattering limit quantitatively account for the measured tunneling conductance. We further demonstrate that the scaling exponent remains quantized to $g=1/3$ throughout the incompressible region of the $\nu=1/3$ plateau, only deviating as the bulk becomes compressible. Finally we examine the impact of weak disorder within the QPC on the tunneling conductance. We demonstrate that the observed chiral Luttinger liquid behavior does not depend sensitively on the presence of non-monotonic features in the zero-bias QPC conductance associated with weak mesoscopic fluctuations. Collectively our measurements and analysis firmly establish the topological robustness of anyon tunneling at the $\nu=1/3$ edge and give strong evidence for the validity of the chiral Luttinger liquid model of edge mode dynamics in the fractional quantum Hall regime. 

The device measured in this experiment consists of a QPC that is an element of a Fabry-P{\'e}rot interferometer \cite{Nakamura2019, Nakamura2020, Nakamura2022, Nakamura2023, Shuang2025}. This device is distinct from the one examined in \cite{guerrerosuarez2025universalanyontunnelingchiral}; it is characterized by an electron density of $n=0.95 \times 10^{11}$~cm$^{-2}$ and a mobility of $\mu = 7.5 \times 10^6$~cm$^2$/Vs in the fully fabricated device configuration. The screening well heterostructure employed here provides a sharp confinement potential at the QPC; this crucial feature is evidently necessary to observe universal chiral Luttinger liquid behavior. To prevent parallel conduction through  the screening wells, we use top and bottom gates to locally deplete the screening wells near the ohmic contacts while the principal quantum well remains galvanically connected to the ohmic leads \cite{EisensteinAPL1990}. Fig.~1a is a false-color scanning electron microscopy image of a device similar to the one measured in this experiment with a schematic of the measurement circuit. Although this device geometry is used to define a Fabry-P{\'e}rot interferometer \cite{Chamon1997}, we only use the QPC highlighted in orange for our tunneling conductance measurements. We also utilize a helper gate, highlighted in green in the Fig.~1a, to locally tune the potential within the QPC.  All other fine gates are held at ground potential. The QPC opening is $300\, \text{nm}$ while the plunger gates are $800\, \text{nm}$ apart. We use standard lock-in techniques with AC excitation of $5\, \mu \text{V}$ and a frequency of $7\, \text{Hz}$. The measurement circuit contains two drain contacts, $D_1$ and $D_2$, to measure both the transmitted current, $I$, and the tunneling current, $I_{\text{tun}}$. Drain contacts are connected to room temperature current amplifiers through cold low-pass RC filters with a resistance 1.2~k$\Omega$ and capacitance 10~nF. 

The differential conductance is defined as $G = \frac{\partial I}{\partial V}\big\vert_{V_{SD}}$ and the tunneling conductance is defined as $G_t = \frac{\partial I_{\text{tun}}}{\partial V}\big\vert_{V_{SD}}$. Fig.~1b displays the magnetotransport through the device. We measure the bulk Hall resistance $R_{xy}$ and the diagonal resistance $R_D$ across the QPC simultaneously. The QPC is biased just past the depletion point of the main quantum well with $V_{QPC} = - 0.1\, \text{V}$. $R_{xy}$ and $R_D$ overlap on all integer plateaus and the primary fractional states including $\nu = 1/3$, indicating the filling factor in the QPC is the same as in the bulk. 

\begin{figure*}[htb!]
    \centering
    \includegraphics[width=\textwidth]{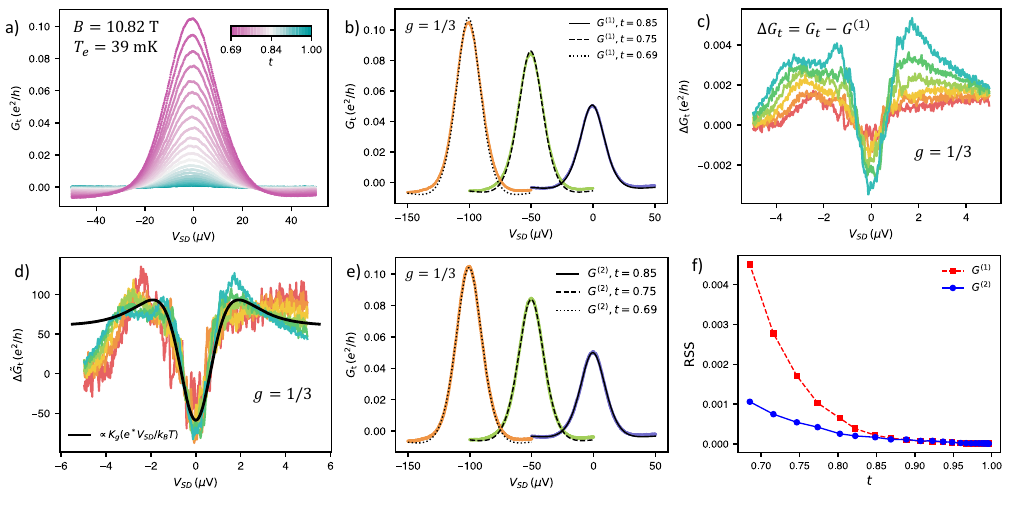}
    \caption{\textbf{(a)} Tunneling conductance versus transmission $t$ as the QPC transmission is varied from $t=0.99$ to $t=0.65$. The magnetic field is fixed at the center of the $\nu = 1/3$ plateau and the electron temperature is fixed at $T_e=0.39~\text{mK}$ as measured via Coulomb blockade thermometry. The color bar indicates the range of transmission through the QPC. \textbf{(b)} Tunneling conductance at $t=0.83, 0.73,$ and $ 0.67$ plotted along with fits to $G^{(1)}$. The plots are staggered by $50\,\mu\mathrm{V}$ for clarity. $G^{(1)}$ accurately captures the data at high transmission but begins to deviate below $t\approx0.8$ as indicated by the dashed and dotted lines. \textbf{(c)} Difference between the data and fit to $G^{(1)}$ from $t=0.81$ to $t=0.67$. $g = 1/3$ while $T_0$ is the only free fitting parameter. $\Delta G_t \approx 0$ for $t >0.8$ but increases in magnitude for lower transmissions. \textbf{(d)} $\Delta G_t$ scaled by $\left(T_0/2\pi T\right)^{4g-4}$ plotted  versus $e^*V_{SD}/k_B T$ for $t=0.81$ to $t=0.67$. The scaled data collapses onto a single curve, signaling that the deviations from the lowest order approximation to the tunneling conductance are described by a universal function of $e^* V_{SD}/k_B T$. The black line displays the next order perturbative contribution to the tunneling conductance, $K_g(e^*V_{SD}/k_B T)$, multiplied by a constant. $K_g(e^*V_{SD}/k_B T)$ accurately captures the functional dependence of the collapse. \textbf{(e)} Tunneling conductance at $t=0.83, 0.73$, and $ 0.67$ plotted with fits to $G^{(2)}$. The plots are staggered by $50\,\mu\mathrm{V}$ for clarity. Fitting to $G^{(2)}$ more accurately captures the data over the full range of transmission and source-drain bias explored here. \textbf{(f)} Residual sum of squares (RSS) for fits to $G^{(1)}$ and $G^{(2)}$ plotted as a function of transmission. The RSS is equal for both functions at high transmission while $G^{(2)}$ performs significantly better as the transmission is lowered.}
    \label{Fig2}
\end{figure*}

Tunneling at the QPC is mediated by anyons in the weak backscattering regime. The weak backscattering regime is defined as $G_t \ll\sigma_{xy}$ \cite{Wen1991, Kane1992}. The tunneling conductance in the weak backscattering limit to lowest order in perturbation theory \cite{Wen1990, Wen1991} is given by:
\begin{align}\label{eq.1}
    G^{(1)} = \frac{e^2}{h}\left(\frac{2 \pi T}{T_0}\right)^{2g-2}F_g\left(\frac{e^* V_{SD}}{k_B T}\right)
\end{align}
where $T$ is the electron temperature, $T_0$ is an effective temperature scale that quantifies the tunnel coupling between edge modes and $V_{SD}$ is the source-drain bias. $e^*$ is the fractional charge of the quasiparticles and $g$ is the scaling exponent. The function $F_g(x)$ is given by
\begin{align}\label{Fg}
    F_g(x) = B\left(g+i\frac{x}{2\pi}, g - i\frac{x}{2\pi}\right)\cosh(x/2)\\\nonumber\times\left\{\pi-2\tanh(x/2)\text{Im}\left[\psi\left(g+i\frac{x}{2\pi}\right)\right]\right\}
\end{align}
here $B$ is the beta function and $\psi$ is the digamma function. The tunneling conductance is a universal function of $T/T_0$ and $e^*V_{SD}/k_B T$. This universal scaling behavior for anyons was first experimentally demonstrated in \cite{guerrerosuarez2025universalanyontunnelingchiral} for specific values of magnetic field and QPC transmission. Here we establish the QPC transmission range over which the tunneling conductance yields a scaling exponent quantized to $g=1/3$ when analyzed within the framework of perturbative calculations. 

Fig.~2a shows the tunneling conductance measured at many distinct values of QPC transmission. The electron temperature is fixed at $T_e = 39$~mK and the transmission varies from $t=0.99$ to $t=0.65$. The electron temperature is measured {\it in operando} using Coulomb blockade thermometry \cite{guerrerosuarez2025universalanyontunnelingchiral}. We fit the tunneling conductance to the lowest order approximation (\ref{eq.1}) with $g=1/3$ and with $T_0$ as the only free parameter (see the Supplementary Information for the evolution of $T_0$ as a function of $t$). The anyon charge $e^*=e/3$ is determined independently by operating the full Fabry-P{\'e}rot interferometer. Representative fits at transmission values of $t=0.83, 0.73$, and 0.67 are shown in Fig.~2b. The fits accurately replicated the data at high transmission but begin to deviate below $t\approx0.8$, as seen most clearly for the data set at $t=0.67$. The difference between the data and the fit, defined as $\Delta G_t \equiv G_t-G^{(1)}$ is plotted in Fig.~2c. We observe that $\Delta G_t \approx 0$ up to $t \approx 0.8$, indicating that the data is well described by the lowest order approximation to the tunneling conductance. At fixed temperature, lower transmission corresponds to a stronger edge coupling and consequently higher rate of quasiparticle tunneling. The observed deviations of the lowest order approximation when compared to the measured data may be attributed to the need to include next order perturbative corrections to the tunneling conductance.

To systematically analyze the deviations between the data and the lowest order perturbative expression for the tunneling conductance, we first rescale $\Delta G_t$ by a factor expected from the next order in a perturbative series in integer powers of $(T/T_0)^{2g-2}$:
\begin{align}
    \Delta \tilde{G}_t = \Delta G_t\left(\frac{T_0}{2\pi T}\right)^{2(2g-2)}
    \;.
\end{align}
Here we examine data from $t=0.81$ to $t=0.67$, where $\Delta G_t$ is non-zero. When $\Delta \tilde{G}_t$ is plotted as a function of $e^*V_{SD}/k_B T$, as shown in Fig.~2d, we observe that the data collapse onto a single functional form. This result indicates that the deviations from the lowest order approximation for the tunneling conductance are described by a universal function of $e^* V_{SD}/k_B T$ that multiplies the next order term in powers of $(T/T_0)^{2g-2}$. 

\begin{figure}[htb!]
    \centering
    \includegraphics[width=\linewidth]{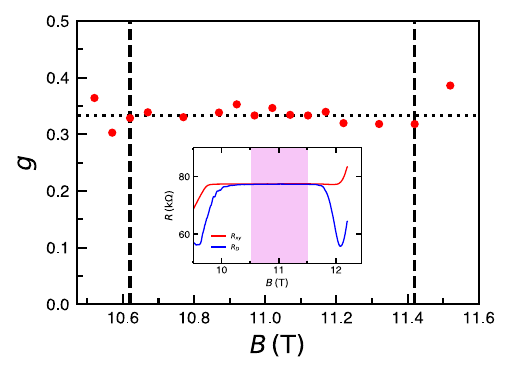}
    \caption{$g$ extracted from fitting the tunneling conductance to $G^{(1)}$ plotted as a function of magnetic field across the $\nu=1/3$ plateau. The magnetic field is varied over a range of $1\, \text{T}$ while the QPC transmission is fixed at $t=0.93$. The vertical dotted lines correspond to the boundaries of the region where $g$ is constant. The inset shows the Hall resistances ($R_{xy}$ and $R_D$) around the $\nu=1/3$ plateau with the translucent box highlighting the region where the tunneling conductance was measured.}
    \label{Fig3}
\end{figure}
\begin{figure*}[htb!]
    \centering
    \includegraphics[width=\textwidth]{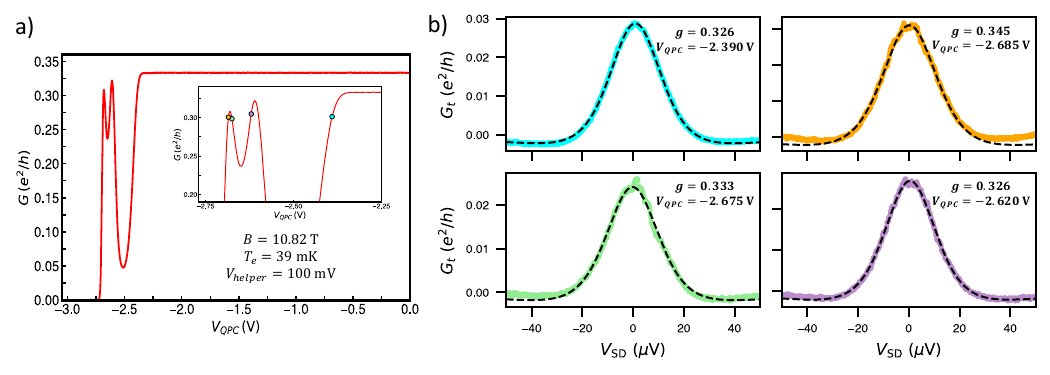}
    \caption{\textbf{(a)} Zero bias QPC conductance at $\nu=1/3$ with $100$~mV applied to the helper gate. The zero bias conductance is quantized to $e^2/3h$ over most of the voltage range and displays a sharp pinch-off. Non-monotonic behavior is evident in the conductance before full pinch-off. The inset highlights these features and the circles display the bias points for the tunneling conductance data shown in Fig.~4b. \textbf{(b)} Tunneling conductance measured at 4 different points along the zero-bias QPC conductance curve. The black lines correspond to the fits to the data. While the voltage applied to the QPC varies, the transmission at all of these points is $t=0.91$.}
    \label{Fig4}
\end{figure*}
The tunneling conductance including the next order correction is given by
\begin{align}\label{eq.2}
    G^{(2)} &= \frac{e^2}{h}\left[\left(\frac{2 \pi T}{T_0}\right)^{2g-2}F_g\left(\frac{e^* V_{SD}}{k_B T}\right)\right.\nonumber\\
    &\hspace{5mm}\left.+\alpha_g\left(\frac{2 \pi T}{T_0}\right)^{4g-4}K_g\left(\frac{e^* V_{SD}}{k_B T}\right)\right]\;,
\end{align}
where the function $K_g(x)$ is computed following Refs.~\cite{Fendley1995,Fendley1996b} and presented in detail in the Supplementary Information. The scaling function $K_g(x)$ captures the line shape of the collapsed data, $\Delta \tilde{G}_t$, when it is rescaled by a factor of $0.5$ and offset vertically by $0.001\,e^2/h$, as shown in Fig.~2d. This observation further indicates that next order corrections included in $G^{(2)}$ account for the deviations of the measured data from $G^{(1)}$. The rescaling of $K_g(x)$ is included through our inclusion of the parameter $\alpha_g$ in the expression for $G^{(2)}$.

Next we directly compare the data with the functional form of $G^{(2)}$ for $t=0.83, 0.73,$ and $0.67$. We fix $g = 1/3$ and $\alpha_g=4g^2\approx0.44$ for all fits, while $T_0$ is a fitting parameter. These fits are plotted in Fig.~2e. Evidently, $G^{(2)}$ better reproduces the exact functional form of the data throughout the range of transmission and $V_{SD}$ studied here. Fig.~2f compares the residual sum of squares (RSS) generated by fitting to $G^{(1)}$ and $G^{(2)}$ as a function of transmission. Although the RSS are equivalent for both functions at high transmission, fits to $G^{(2)}$ are significantly better as the transmission is reduced below $t=0.8$. This clearly demonstrates that by lowering $t$ we are leaving the weak backscattering regime and that the inclusion of the next order correction is necessary to capture detailed features in the data.

We study the magnetic field dependence of the extracted tunneling exponent across the $\nu = 1/3$ plateau. We focus on the regime of weak backscattering with $t=0.93$ where the $G^{(1)}$ is expected to be a good approximation to the tunneling conductance. Fig.~3 displays $g$ extracted by fitting the tunneling conductance to equation (1) at several values of magnetic field. The transmission at all points is fixed to $t=0.93$. It is important to note that $g$ is not fixed to $g=1/3$ but is extracted from the fit for these data sets investigating the magnetic field dependence of anyon tunneling using the same approach as used in Ref.~{\cite{guerrerosuarez2025universalanyontunnelingchiral}}. The inset highlights the region on the plateau over which we measure the tunneling conductance. We find that $g$ remains constant over a range of $B\sim \, 1\text{T}$ before it begins to deviate. Outside of this range, the quantum Hall state starts to become compressible, consistent with the observed deviation in the extracted value of $g$. The average value of $g$ within the incompressible region is $g = 0.333 \pm 0.001$, in excellent agreement with the theoretical expectation of $g = 1/3$. This data demonstrates the preservation of quantization of $g$ within the $\nu=1/3$ incompressible quantum Hall state, consistent with the theory of the chiral Luttinger liquid. 

Finally, we study the impact of mesoscopic disorder within the QPC, as reflected in non-monotonic features in the linear conductance of the QPC immediately prior to pinch-off, on the tunneling conductance and scaling exponent. Fig.~4a shows the zero bias conductance of the QPC where we have intentionally used the helper gate to accentuate non-monotonic behavior. We apply a voltage of $100\, \text{mV}$ for the measurements shown in Fig.~4. As we vary the QPC voltage while keeping the helper gate fixed at $100$~mV, the linear conductance is quantized to $e^2/3h$ over most of the voltage range before showing a steep drop to zero. Immediately before pinch off, non-monotonic behavior is evident in the conductance. The tunneling conductance is measured at four different $V_{QPC}$ points, all of which yield transmission of $t=0.91$ as indicated by the orange, blue, magenta and green dots in Fig.~4a. The tunneling conductance data and fits are shown in Fig.~4b. We observe that the scaling exponent remains close to $g = 1/3$ at all of these points. The appearance of weak disorder within the QPC does not appear to alter the scaling exponent describing the tunneling of anyons between chiral edge modes. As the $\nu=1/3$ bulk fractional quantum Hall state is associated with a single chiral edge mode, we do not expect weak disorder to change the scaling exponent~\cite{Haldane1995}. Of course for more complex states with multimode edges, such $\nu=2/3$ and $\nu=2/5$, disorder may be expected to alter the edge modes~\cite{KaneFisher1995, KaneFisherPolchinski}. Multimode edge states will be explored in future work.

Our experiments reveal robust chiral Luttinger liquid behavior for the $\nu = 1/3$ fractional quantum Hall edge mode and demonstrate the relative insensitivity of the scaling exponent $g$ to perturbations that result in non-monotonic features in the zero-bias QPC conductance near pinch-off. This experimental demonstration of robustness against multiple perturbations substantiates the chiral Luttinger liquid theory of the fractional quantum Hall edge mode at $\nu=1/3$ in a manner not previously possible, and validates the utility of the scaling exponent $g$ for the identification of topological order in the bulk. Our analysis of tunneling conductance beyond lowest-order perturbation reveals the importance of correlations among anyon tunneling events as backscattering increases.
The robustness of the chiral Luttinger liquid behavior observed in our study establishes that edge tunneling experiments -- with sharp edge confinement provided by the screening-well heterostructure design -- can be used in combination with Fabry-P{\'e}rot interferometry to characterize the topological order of the bulk state.
\newline

\begin{acknowledgments}
This research is sponsored by U.S. Department of Energy, Office of Science, Office of Basic Energy Sciences, under the award numbers DE-SC0020138 for the experimental work and DE-FG02-06ER46316 for theory. We thank James Nakamura for device fabrication, and Paul Fendley for helpful and insightful discussions on the calculation of higher-order corrections to the tunneling conductance.
\end{acknowledgments}

\bibliography{apssamp}

\end{document}


\title{Supplementary Information for ``Topological Robustness of Anyon Tunneling at $\nu = 1/3$"}

\author{Adithya Suresh}
\affiliation{Department of Physics and Astronomy, Purdue University, West Lafayette, IN, 47907}
\author{Ramon Guerrero-Suarez}
\affiliation{Department of Physics and Astronomy, Purdue University, West Lafayette, IN, 47907}
\author{Tanmay Maiti}
\affiliation{Department of Physics and Astronomy, Purdue University, West Lafayette, IN, 47907}
\author{Shuang Liang}
\affiliation{Department of Physics and Astronomy, Purdue University, West Lafayette, IN, 47907}
\author{Geoffrey Gardner}
\affiliation{Microsoft Quantum, West Lafayette, IN, 47907}
\author{Claudio Chamon}
\affiliation{Department of Physics and Astronomy, Purdue University, West Lafayette, IN, 47907}
\affiliation{Purdue Quantum Science and Engineering Institute, Purdue University, West Lafayette, IN, 47907}
\author{Michael Manfra}
\affiliation{Department of Physics and Astronomy, Purdue University, West Lafayette, IN, 47907}
\affiliation{Microsoft Quantum, West Lafayette, IN, 47907}
\affiliation{Purdue Quantum Science and Engineering Institute, Purdue University, West Lafayette, IN, 47907}
\affiliation{Elmore Family School of Electrical and Computer Engineering, Purdue University, West Lafayette, IN, 47907}
\affiliation{School of Materials Engineering, Purdue University, West Lafayette, IN, 47907}

\maketitle

\makeatletter
\renewcommand{\thefigure}{S\arabic{figure}}
\makeatother
\setcounter{figure}{0}

\section{Calculation of excess tunneling current}
Our measurement circuit consists of a source contact where we apply a voltage excitation $V_S$ and two drains, $D_1$ and $D_2$, used for measuring the transmitted current $I$ and tunneling current $I_{\text{tun}}$ respectively. $D_1$ and $D_2$ are connected to room temperature current amplifiers through cold RC filters with component values 1k$\Omega$ and 10nF followed by a $200\,\Omega$ line resistance. In general, $I \neq I_{\text{tun}}$ which implies the two drains are not at the same potential. This causes a current to flow from $D_1$ to $D_2$ even when there is no backscattering from the QPC. This excess current must be subtracted from the tunneling current measured at $D_2$. Calculation of the excess current using Landauer-Büttiker formalism is shown below. Current conservation implies:
\begin{align}
    \frac{e^2}{3h}V_S &= I_S \\[5 pt]
    -t\,I_S+\frac{e^2}{3h}V_{D_1}&= -I_{D_1}\\[5 pt]
    -t\frac{e^2}{3h}V_{D_1}-(1-t)I_S+\frac{e^2}{3h}V_{D_2}& = -I_{D_2}
\end{align}
where $t$ is the transmission of the QPC. These equations are straightforward to solve noting that $V_{D_1} = I_{D_1}R$ and $V_{D_2} = I_{D_2}R$, where $R = 1.2$\,k$\Omega$ is the sum of the line and filter resistances. We are interested in $I_{D_2}$, which is given by
\begin{align}\label{Id2sol}
    I_{D_2} = t\,I_{D_1}\left(\frac{R}{R+R_H}\right)+(1-t)I_S
\end{align}
where $R_H$ is the Hall resistance. At $\nu = 1/3$, $R_H \approx 77.436\, \text{k}\Omega$. The second term is the backscattered current from the QPC and the first term is the excess current. $I_{D_1}$ is measured to be $63.6\,\text{pA}$ in our circuit. Substituting these values into \eqref{Id2sol}, for $t=1$, gives $I_{D_2} \approx 0.97\,\text{pA}$.
\newline

\section{Transmission dependence of $T_0$}
$T_0$ is extracted from the fits to the tunneling conductance data and is a measure of the coupling of the counterpropagating edge modes across the QPC. $T_0$ is expected to increase monotonically as the transmission $t$ is reduced. 
\begin{figure}[H]
    \centering
    \includegraphics[width=\linewidth]{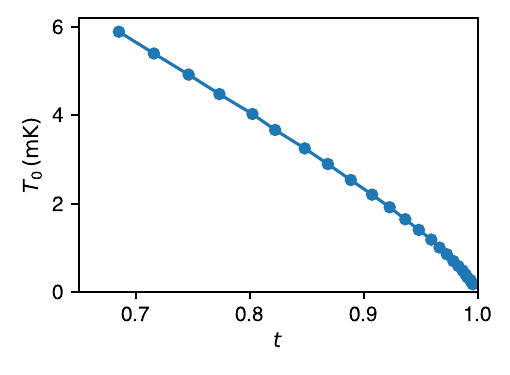}
    \caption{Extracted $T_0$ as a function of transmission $t$ for fits to $G^{(2)}$. $T_0$ increases monotonically with decreasing transmission.}
    \label{FigS1}
\end{figure}

\begin{figure*}[!htb]
    \centering
    \includegraphics[width=\textwidth]{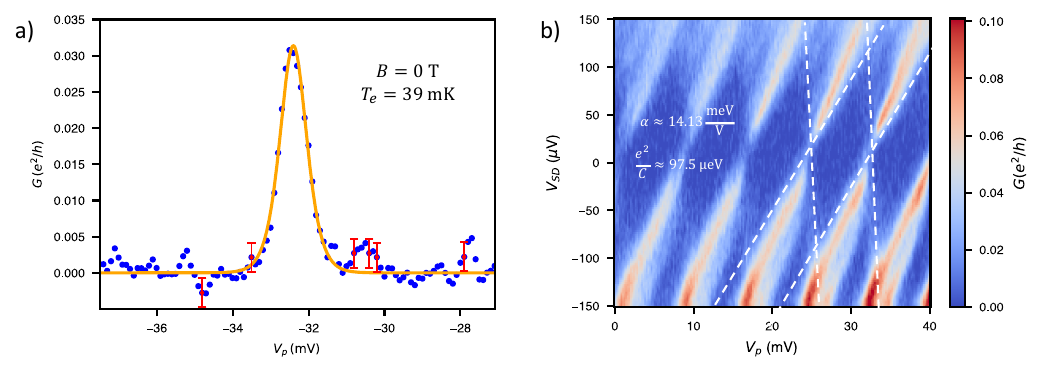}
    \caption{\textbf{(a)} Zero bias conductance as a function of plunger gate voltage in the Coulomb blockade regime. Representative error bars are shown for some of the points near $G = 0$. The extracted electron temperature is $T_e = 39\,\text{mK}$. \textbf{(b)} Two-dimensional color map of conductance through the quantum dot as a function of source-drain bias and plunger gate voltage. The white dotted lines highlight the approximate range of the Coulomb diamond.}
    \label{FigS2}
\end{figure*}
\section{Coulomb blockade thermometry}
The QPC used for tunneling conductance measurements is part of a Fabry-P{\'e}rot interferometer. We operate the interferometer as a quantum dot by putting the QPCs into the tunneling regime. The zero bias conductance, as shown in Fig.~S2a, is then related to the plunger gate voltage by $G \propto \cosh^{-2}\left(\alpha V_p/2k_B T_e\right)$, where $\alpha$ is the lever arm of the plunger gate to the quantum dot. The lever arm is extracted from analysis of the Coulomb diamonds shown in Fig.~S2b; here $\alpha \approx 14.13\,\text{meV}/\text{V}$. We fit the data to extract the electron temperature $T_e$. In this instance, we extract an electron temperature of $T_e=39\,\text{mK}$.

\begin{widetext}
\section{Calculation of the tunneling conductance to fourth order in the
quasiparticle tunneling amplitude}

The QPC brings left and right propagating edge modes into close proximity. If the region between the left and right moving edges contains an FQHE state in which fractional quasiparticles are supported, then quasiparticles can tunnel between the edges. In contrast, when the FQHE state is completely removed from the region in between the edges, only integer valued charges, i.e. electrons, can tunnel. The experiments described in this paper are carried out in a regime of intermediate-to-high transmissions consistent with weak quasiparticle tunneling, in which tunneling is described by the operator~\cite{Wen1990, Wen1991}
\begin{align}
    H_{\rm tun} = \Gamma e^{i\omega_0 t} \;\Psi_L^\dagger\,\Psi_R + H.c.,
\end{align}
where
\begin{align}
    \Psi_{L,R} \sim e^{\mp i \sqrt{g}\phi_{L,R}}
    \;,
\end{align}
with $\phi_{L,R}$ the bosonic fields from the chiral Luttinger liquid theory, $g=\nu$, and $\omega_0=\frac{\nu e V}{\hbar}$. The dynamics of these fields is governed by the Lagrangian:
\begin{align}
    {\cal L}_{L,R} = \frac{1}{4\pi} \partial_x \phi_{L,R} \left(\mp\partial_t - v\partial_x \right)\phi_{L,R}
    \;,
\end{align}
with equal-time commutation relation $[\phi_{L,R}(t,x), \phi_{L,R}(t,y)] = \mp i\pi\;\text{sgn}(x-y)$. The charge density along the edges is given by $\rho_{L,R}=\partial_x\phi_{L,R}$.

The full Lagrangian, including both edges and the tunneling term, can be recast as
\begin{align}
    {\cal L} = \frac{1}{8\pi} \left[(\partial_t \phi)^2-v^2 (\partial_x \phi)^2\right] - \Gamma\delta(x) e^{i\omega_0 t} \;e^{i\phi} +H.c.
    \;,
    \label{eq:Lagrangian}
\end{align}
where $\phi=\phi_L+\phi_R$. Notice that the tunneling term is active only at a single point in space, the origin, as can be read from the $\delta$-function. The tunneling conductance can be computed in perturbation theory in the tunneling amplitude $\Gamma$, by developing the perturbative expansion of the partition function in terms of a gas of positive and negative particles (associated with insertions of the vertex operators $e^{\pm i\phi}$)
interacting with a two-dimensional (logarithmic) Coulomb potential that originates from the $\phi$ correlations due to the free part of the Lagrangian~\eqref{eq:Lagrangian}. The fugacities of the positive and negative charges are proportional to the
tunneling amplitude and its complex conjugate, $\Gamma$ 
and $\Gamma^{\ast}$. Charge neutrality forces the powers of
$\Gamma$ and $\Gamma^*$ to be equal. This expansion can be developed using a Keldysh contour, to account for the non-equilibrium bias voltage~\cite{ChamonFreedWen1995}. It can also be related to the partition function of the boundary sine-Gordon model (the charges are inserted at 
different times but always at $x=0$, because of the $\delta$-function, in contrast to the sine-Gordon model whose 
counterpart action does not contain the $\delta$-function at the origin.) The partition function 
for the boundary sine-Gordon model can be expressed in terms of 
Jack polynomials, as formulated by Fendley, Lesage, and Saleur in 
Refs.~\cite{Fendley1995,Fendley1996b}. The calculation below follows directly their approach in Ref.~\cite{Fendley1996b}.

The integrals over
charge positions can be computed by expanding the integrands in terms of Jack
polynomials and using their orthogonality properties. The calculation can be
extended to the non-equilibrium case in which a voltage bias is applied.
Ref.~\cite{Fendley1996b} finds a series
\begin{align}
  Z_{BSG} (x_{\tmop{BSG}}, p)  = 1 + \sum_{n =
    1}^{\infty} (x_{\tmop{BSG}} )^{2 n}  I_{2 n} (p) \;,
\end{align}
where $(x_{\tmop{BSG}} )^2 \propto | \Gamma |^2 (2 \pi T)^{2 (g - 1)}$ and $p
= i \frac{g V}{2 \pi T}$, with $T$ the temperature, and $V$ the applied
voltage. The coefficients $I_{2 n} (p)$ in this power series (in
$x_{\tmop{BSG}} $) expansion are
\begin{align}
  I_{2 n} (p) = \frac{1}{\Gamma (g)^{2 n}} \sum_{\{ m_1, \ldots, m_n
    \}} \prod_{i = 1}^n \frac{\Gamma (m_i + g (n - i +
    1)) \hspace{0.27em} \Gamma (p + m_i + g (n - i +
    1))}{\Gamma  (m_i + 1 + g (n - i)) 
    \Gamma  (p + m_i + 1 + g (n - i))},
\end{align}
with $g$ the scaling exponent and the set $m_1 \geqslant m_2
\geqslant \cdots \geqslant m_n$ a partition of integers according to
Young tableaux with $n$ rows.

The conductance was conjectured in Ref.~\cite{Fendley1995} and checked against Bethe Ansatz
results in Refs.~\cite{Fendley1995PRL,Fendley1995PRB} to be \ \
\begin{eqnarray}
  G \left( x_{\tmop{BSG}} , \frac{V}{2 T} \right)
  & = &
  g
  - ig  \pi \frac{x_{\tmop{BSG}} }{2}  \frac{\partial}{\partial (V / 2 T)} 
  \frac{\partial}{\partial x_{\tmop{BSG}} } \ln \left( \frac{Z_{\tmop{BSG}}
  \hspace{-0.17em} \left( x_{\tmop{BSG}} , \tfrac{i gV}{2 \pi T}
  \right)}{Z_{\tmop{BSG}}  \left( x_{\tmop{BSG}} , - \tfrac{i
      gV}{2 \pi T} \right)} \right)
  \nonumber\\
  & = &
  g + g^2  \frac{x_{\tmop{BSG}} }{2}  \frac{\partial}{\partial_  p} 
  \frac{\partial}{\partial x_{\tmop{BSG}} } \ln \left( \frac{Z_{\tmop{BSG}} 
   (x_{\tmop{BSG}} , p)}{Z_{\tmop{BSG}}  
  (x_{\tmop{BSG}} , - p)} \right) \Huge{\left| \large{_{p = i \frac{g V}{2 \pi
  T}} \Huge{}} \right.},
\end{eqnarray}
in units of $\frac{e^2}{h} .$ (We work in units of $e = \hbar = 1$.)

Here we shall focus on the tunneling conductance $G_t = g - G$ up to fourth
order in $x_{\tmop{BSG}} $, so we expand the partition function to that order:
\begin{align}
  \begin{array}{lll}
     \ln Z_{\tmop{BSG}} (x_{\tmop{BSG}} , p) & = & \left( {x_{\tmop{BSG}} }  
     \right)^2 I_2 (p) + \left( {x_{\tmop{BSG}} }   \right)^4 I_4 (p) -
     \frac{1}{2}  \left( {x_{\tmop{BSG}} }   \right)^4  [I_2 (p)]^2 + \cdots
  \end{array}
\end{align}
The conductance is then given by
\begin{subequations}
\begin{align}
  G_t = G^{(1)}_t + G^{(2, 2)}_t + G^{(2, 4)}_t \;,
\end{align}
where
\begin{eqnarray}
  G^{(1)}_t & = & - 2 g^2  \left( {x_{\tmop{BSG}} }   \right)^2 \tmop{Re}
  [\partial_p I_2 (p)] \left|_{p = i \frac{g V}{2 \pi T}}
  \right._{\large{\Huge{}}} \quad,\\
  G^{(2, 2)}_t & = & + 4 g^2  \left( {x_{\tmop{BSG}} }   \right)^4 \tmop{Re}
  [I_2 (p) \partial_p I_2 (p)] \left|_{p = i \frac{g V}{2 \pi T}}
  \right._{\large{\Huge{}}} \quad,\\
  G^{(2, 4)}_t & = & - 4 g^2  \left( {x_{\tmop{BSG}} }   \right)^4 \tmop{Re}
  [\partial_p I_4 (p)] \left|_{p = i \frac{g V}{2 \pi T}}
  \right._{\large{\Huge{}}} \quad .
\end{eqnarray}
\end{subequations}
We proceed to compute the three contributions in each line of the equation
above.

To do that, we need
\begin{eqnarray}
  I_2 (p)
  & = &
  \frac{1}{\Gamma (g)^2} \sum_{m_1 = 0}^{\infty}
  \frac{\Gamma (m_1 + g) \Gamma (p + m_1 + g)}{\Gamma (m_1 + 1)
    \Gamma (p + m_1 + 1)}\nonumber\\
  & = &
  \frac{1}{2 \pi}\; B (g + p, g - p)
  \; \frac{\sin [\pi (g - p)]}{\cos (\pi g)}
   \end{eqnarray}
and
\begin{eqnarray}
  I_4 (p)
  & = &
  \frac{1}{\Gamma (g)^4}  \sum_{m_1 = 0}^{\infty}
  \sum_{m_2 = 0}^{m_1} \frac{\Gamma (m_1 + 2 g) 
    \Gamma (p + m_1 + 2 g)}{\Gamma (m_1 + 1 + g) 
    \Gamma (p + m_1 + 1 + g)}
  \frac{\Gamma (m_2 + g)  \Gamma (p + m_2 + g)}{\Gamma (m_2 + 1) 
  \Gamma (p + m_2 + 1)} .
\end{eqnarray}
Using these expressions, one can compute each of the corrections
$G^{(1)}_t, G^{(2, 2)}_t$ and $G^{(2, 4)}_t$. The first two can be
more simply computed, and yield
\begin{subequations}
\begin{eqnarray}
  G^{(1)}_t (x = g V / T)
  & = &
  \left( {x_{\tmop{BSG}} }   \right)^2 
  \frac{g^2}{\pi} F_g (x),\\
  G^{(2, 2)}_t (x = g V / T) & = & - \left( {x_{\tmop{BSG}} }   \right)^4 
  \frac{g^2 \; \tan (\pi g)}{\pi^2} H_g (x),\\
  G^{(2, 4)}_t (x = g V / T) & = & - \left( {x_{\tmop{BSG}} }   \right)^4 2
  g^2 Q_g (x),
\end{eqnarray}
where
\begin{align}
  F_g (x) = B \left( g + \tfrac{i x}{2 \pi}, g - \tfrac{i x}{2 \pi} \right)
    \left\{ - 2 \; \tmop{Im} \psi \left( g + \tfrac{i x}{2 \pi}
   \right) \sinh \left( \frac{x}{2} \right) + \pi \; \cosh \left( \frac{x}{2}
   \right) \right\} \;,
\end{align}
\begin{align}
  H_g (x) = \left[ B \left( g + \tfrac{i x}{2 \pi}, g - \tfrac{i x}{2 \pi}
   \right) \right]^2  \left\{ - 2 \; \tmop{Im} \psi \left( g +
  \tfrac{i x}{2 \pi} \right) \sinh (x) + \pi \; \cosh (x) \right\} \;,
\end{align}
and
\begin{eqnarray}
  Q_g (x) & = & 2 \tmop{Re} \huge[\frac{1}{\Gamma (g)^4}  \sum_{m_1 = 0}^{\infty}
  \sum_{m_2 = 0}^{m_1} \frac{\Gamma (m_1 + 2 g) 
    \Gamma (p + m_1 + 2 g)}{\Gamma (m_1 + 1 + g) 
    \Gamma (p + m_1 + 1 + g)} \frac{\Gamma (m_2 + g)
    \Gamma (p + m_2 + g)}{\Gamma (m_2 + 1) 
    \Gamma (p + m_2 + 1)}
  \nonumber\\
  && \times
  [\psi (2 g + m_1 + p) + \psi (g + m_2 + p) - \psi (1 + g + m_1 + p) -
  \psi (1 + m_2 + p)] \Huge|_{p = i \frac{x}{2 \pi} x} \huge]
\end{eqnarray}
\end{subequations}
The last expression can be computed numerically by defining a quantity
$Q_g (x ; \mathcal{N}) $ in which we truncate the sum over $m_1$ to
$\sum_{m_1 = 0}^{\mathcal{N}}$. We compute the sum up to $\mathcal{N} = 20000$. Then we use the fact that the sum converges as $m_1^{-4/3}$ as $m_1 \rightarrow \infty$, to approximate the tail for $m_1 > \mathcal{N}$. 

Finally, identifying \ $\left( {x_{\tmop{BSG}} }   \right)^2  \frac{g^2}{\pi}
= \left( \frac{2 \pi T}{T_0} \right)^{2 (g - 1)}$, we can write
\begin{subequations}
\begin{align}
  G_t \left( x = \frac{g V}{T}, \frac{T}{T_0}  \right) = \left( \frac{2 \pi
   T}{T_0} \right)^{2 (g - 1)} F_g (x) - \left( \frac{2 \pi T}{T_0} \right)^{4
   (g - 1)} \left[ \frac{\; \tan (\pi g)}{g^2} H_g (x) + \frac{2 \pi^2}{g^2}
    Q_g (x) \right] \; .
\end{align}
For the discussion below, we combine the two contributions to fourth order into a single function $K_g(x)$ given by
\begin{align}
    K_g(x)=-\left[ \frac{\; \tan (\pi g)}{g^2} H_g (x) + \frac{2 \pi^2}{g^2}
    Q_g (x) \right] \;.
\end{align}
\end{subequations}

These results yield the formula for $G^{(2)}$ used in the main text to compare with the data, namely,
\begin{align}
    G^{(2)} &= \frac{e^2}{h}\left[\left(\frac{2 \pi T}{T_0}\right)^{2g-2}F_g\left(\frac{e^* V_{SD}}{k_B T}\right)\right.\nonumber\\
    &\hspace{5mm}\left.+\alpha_g\left(\frac{2 \pi T}{T_0}\right)^{4g-4}K_g\left(\frac{e^* V_{SD}}{k_B T}\right)\right]\;,
\end{align}
where we include a scaling factor $\alpha_g$ discussed in the main text.

\end{widetext}



\bibliography{apssamp}